\documentclass{article}
\usepackage{dcase2026_techrep,amsmath,graphicx,url,times,booktabs,tabularx,hyperref}
\hypersetup{hidelinks}
\usepackage{algorithm}
\usepackage{algpseudocode}
\usepackage{amssymb}

\newcommand{\vect}[1]{\boldsymbol{#1}}
\newcommand{\method}{CARE-DPP}

\title{DETERMINANTAL POINT PROCESS SAMPLING FOR BIOACOUSTIC ACTIVE LEARNING}

\name{Hugo Magaldi$^{1*}$,
       Gabriel Dubus$^{1}$,
       }
\address{Eco-Anthropologie, Mus\'eum National d'Histoire Naturelle, UMR7206, CNRS, Paris, France\\
         hugo.magaldi@mnhn.fr}

\begin{document}
\ninept
\maketitle
\begin{sloppy}

\begin{abstract}
Eco-acoustic monitoring generates vast volumes of audio data, making active learning a promising approach for reducing annotation effort while efficiently training reliable biodiversity classifiers. This report presents \method, a batch active-learning acquisition method submitted to BioDCASE Active Learning for Bioacoustics 2026 challenge. The method combines class-balanced predictive uncertainty with embedding-space novelty, while a determinantal point process (DPP) objective selects a high-quality and non-redundant acquisition batch. The uncertainty-novelty balance is annealed over the annotation budget: early cycles emphasize geometric coverage, whereas later cycles increasingly exploit classifier uncertainty. To mitigate unreliable early scores, the DPP candidate pool mixes top-quality candidates with a decreasing proportion of random exploration. An adaptive acquisition schedule uses smaller batches early and larger batches later. Evaluated over five repeats on the BirdSet HSN, POW and UHH subsets and on ATBFL, \method\ obtains a mean development AULC of 0.50 for macro mAP, compared with 0.46 for the official CoreSet baseline. Ablations identify DPP batch diversification and the adaptive acquisition schedule as the largest contributors.
\end{abstract}


\section{Introduction}
Passive acoustic monitoring can produce far more recordings than experts can annotate. Pool-based active learning aims at addressing this imbalance by repeatedly selecting a small set of unlabeled samples whose annotation is expected to improve a model most efficiently \cite{ren_survey_2022,stowell_2022}. The BioDCASE Active Learning for Bioacoustics 2026 task standardizes this setting across terrestrial and marine bioacoustics by providing fixed Perch v2 embeddings and a fixed classification pipeline, while restricting system design to the sampling function and acquisition-batch schedule, and ranking methods based on the area under the macro-mAP learning curve (AULC) up to a fixed annotation budget \cite{biodcase2026task4,mcewen_baseal}.

A successful batch acquisition rule must balance several competing goals. Uncertainty sampling targets decision-boundary examples but can be unreliable when few labels are available and can repeatedly select similar samples. Pure geometric coverage, such as CoreSet \cite{sener_savarese_2018}, is stable early but ignores the evolving classifier and may select acoustically novel samples that are not label-informative. Moreover, macro mAP rewards performance on every class equally, whereas ordinary multilabel uncertainty may be dominated by frequently observed classes.

We propose \method (Class-balanced Annealed Random-Exploration DPP), a method with four components: (i) class-balanced multilabel uncertainty, (ii) cosine novelty relative to the labeled set, (iii) annealed exploration-exploitation weights and candidate-pool exploration, and (iv) DPP-based batch diversification.

Determinantal point processes, originally introduced as repulsive point-process models~\cite{macchi1975fermion}, assign higher probability to subsets whose feature vectors span a large volume~\cite{kulesza_taskar_2012}. Similarity between vector directions promotes diversity, while externally defined quality scores scale their magnitudes. This makes DPPs well suited to batch active learning, where informative samples should be selected without introducing redundancy within each annotation batch~\cite{biyik2019batch}.

\section{Task and Data}
The BaseAL loop starts from a randomly initialized multilabel classification head. At each cycle, the sampler receives current predictions, Perch v2 embeddings, and the currently labeled indices; selected labels are revealed by an oracle and the head is retrained. Perch v2 is a multi-taxa bioacoustic representation model designed for transfer learning \cite{perch2}. The fixed task learning rate is $10^{-3}$ and the model is trained for 10 epochs per cycle.

The development benchmark contains three BirdSet subsets and one aggregated ATBFL dataset. BirdSet is a large-scale avian benchmark \cite{rauch_birdset_2025,rauch_2026_19191603}; ATBFL contains Antarctic blue- and fin-whale call annotations across deployments \cite{kurinchi_vendhan_2026_19133112}. Table~\ref{tab:data} summarizes the development pools used in this work.

\begin{table}[t]
\centering
\small
\setlength{\tabcolsep}{4pt}
\begin{tabular}{lrrr}
\toprule
\textbf{Dataset} & \textbf{Train seg.} & \textbf{Classes} & \textbf{Labels/sample} \\
\midrule
BirdSet HSN & 6,600 & 19 & 0.524 \\
BirdSet POW & 2,280 & 41 & 2.833 \\
BirdSet UHH & 18,319 & 25 & 1.058 \\
ATBFL (all deployments) & 9,086 & 7 & 2.267 \\
\bottomrule
\end{tabular}
\caption{Development-pool summary. ATBFL statistics aggregate its site-year deployments.}
\label{tab:data}
\end{table}

\section{Method}
Let $L_t$ and $U_t$ be the labeled and unlabeled sets at cycle $t$, and let $B_t$ be the requested acquisition size. Each sample has a normalized embedding $\vect{x}_i$ and multilabel probabilities $p_{ic}$ for classes $c=1,\ldots,C$. The scalar weights and thresholds used below were chosen from simple design principles - favoring coverage early, uncertainty later, and diversity within each batch - and then fixed after a coarse validation sweep.

\subsection{Class-balanced uncertainty}
For each class, binary entropy is
\begin{equation}
 h_{ic}=-\frac{p_{ic}\log p_{ic}+(1-p_{ic})\log(1-p_{ic})}{\log 2}.
\end{equation}
Let $n_c$ be the number of currently labeled positives for class $c$. Classes receive weight
\begin{equation}
 a_c \propto (n_c+1)^{-1/2},
\end{equation}
with weights clipped for stability. The class-balanced entropy is blended with standard entropy to bias acquisition toward underrepresented classes while avoiding overreaction to noisy rare-class predictions:
\begin{equation}
 u_i = 0.75\frac{\sum_c a_c h_{ic}}{\sum_c a_c}+0.25\frac{1}{C}\sum_c h_{ic}.
\end{equation}
Only labels already revealed by the active-learning loop are used.

\subsection{Embedding novelty and annealed quality}
Cosine novelty measures distance from the current labeled set:
\begin{equation}
 v_i = 1-\max_{j\in L_t}\vect{x}_i^\top\vect{x}_j.
\end{equation}
Both $u_i$ and $v_i$ are min-max normalized over $U_t$. With normalized budget progress $\tau_t=|L_t|/500$, raw weights evolve as
\begin{equation}
 \bar{w}_u=0.25+0.40\tau_t, \qquad
 \bar{w}_v=0.65-0.40\tau_t.
\end{equation}
They are renormalized to sum to one, and define candidate quality
\begin{equation}
 q_i=w_u u_i+w_v v_i.
\end{equation}
Thus early cycles prioritize coverage, while later cycles increasingly exploit the trained classifier.

\subsection{Candidate exploration and DPP selection}
Applying DPP selection to the complete pool is computationally expensive. We therefore construct a candidate pool of size
\begin{equation}
 M_t=\min\{|U_t|,\max(30B_t,1500)\}.
\end{equation}
The pool contains the highest-quality samples plus a random exploration fraction $\rho_t$:
\begin{equation}
 \rho_t=\begin{cases}
 0.40,& |L_t|<100,\\
 0.25,& 100\leq |L_t|<300,\\
 0.15,& |L_t|\geq 300.
 \end{cases}
\end{equation}
Random candidates broaden the region visible to DPP when model-derived scores are unreliable. We decrease their proportion as the model becomes informative.

For candidate $i$, define $\tilde q_i=0.05+q_i$ and $\vect{z}_i=\tilde q_i\vect{x}_i$. The positive semidefinite DPP kernel is
\begin{equation}
 K_{ij}=\vect{z}_i^\top\vect{z}_j.
\end{equation}

At each cycle $t$, we select a batch $S_t$ by greedily maximizing $\log \det(K_{S_t}+10^{-6} I)$ using pivoted Cholesky ~\cite{harbrecht2012pivoted}, where $K_{S_t}$ is the submatrix of the DPP kernel restricted to the selected samples and $10^{-6} I$ is a regularization term. The determinant rewards individually high-quality samples while penalizing mutually similar samples \cite{kulesza_taskar_2012}. We use the small floor $0.05$ in $\tilde q_i$ to prevent low-scored but potentially diverse candidates from being completely suppressed and improve numerical stability of the log-determinant selection.

\subsection{Adaptive acquisition schedule}
The acquisition batch size is
\begin{equation}
 B_t=\begin{cases}
 25,& |L_t|<100,\\
 50,& 100\leq |L_t|<300,\\
 75,& |L_t|\geq300,
 \end{cases}
\end{equation}
with the final batch truncated at the budget. Smaller early batches allow frequent model updates when each annotation is most influential, while larger late batches reduce repeated retraining.

\begin{algorithm}[t]
\caption{\method\ acquisition at cycle $t$}
\label{alg:dpp}
\begin{algorithmic}[1]
\State Compute class-balanced uncertainty $u_i$ and novelty $v_i$
\State Anneal weights and compute $q_i=w_u u_i+w_v v_i$
\State Build a candidate pool from top $q_i$ scores and random exploration
\State Form $K_{ij}=\tilde q_i \tilde q_j\vect{x}_i^\top\vect{x}_j$
\State Greedily select $B_t$ pivots maximizing incremental log determinant
\State Query labels, update $L_t$, and retrain for 10 epochs
\end{algorithmic}
\end{algorithm}

\section{Experiments and Results}
All results use the same maximum budget of 500, the fixed task learning rate and training epochs, five independent repeats, and AULC as the ranking metric. Submitted YAML files contain the per-cycle means and standard deviations.

\subsection{Main results}
Table~\ref{tab:overall} compares \method\ with official aggregate baselines. The proposed sampler reaches 0.50 mean AULC and 0.59 mean final macro mAP. Per-dataset behavior is analyzed through the ablation study in Table~\ref{tab:ablation}, whose first row gives the full method results on each dataset.

\begin{table}[t]
\centering
\small
\setlength{\tabcolsep}{5pt}
\begin{tabular}{lc}
\toprule
\textbf{Method} & \textbf{Mean AULC} \\
\midrule
\method\ (ours) & \textbf{0.5017} \\
CoreSet (official baseline) & 0.4600 \\
TypiClust (official baseline) & 0.4230 \\
Margin (official baseline) & 0.3990 \\
Random (official baseline) & 0.3900 \\
\bottomrule
\end{tabular}
\caption{Development AULC averaged across datasets. Official baseline values are reported by the task organizers \cite{biodcase2026task4}.}
\label{tab:overall}
\end{table}

\subsection{Ablation study}
Table~\ref{tab:ablation} and Fig.~\ref{fig:ablation} summarize the ablation study. The table reports AULC separately for each dataset, revealing that different components contribute unevenly across domains. Removing DPP causes the largest loss, especially on HSN and UHH, confirming that batch-level non-redundancy is central. Replacing the adaptive acquisition schedule with fixed batches of 50 also produces a substantial drop. Annealing, class balancing and the strong exploration schedule provide smaller but complementary effects.

\begin{table}[t]
\centering
\small
\setlength{\tabcolsep}{2.8pt}
\begin{tabular}{lccccc}
\toprule
\textbf{Variant} & \textbf{ATBFL} & \textbf{HSN} & \textbf{POW} & \textbf{UHH} & \textbf{Mean} \\
\midrule
\method\ (full) & \textbf{0.4652} & 0.6080 & 0.5002 & 0.4335 & \textbf{0.5017} \\
Fixed exploration fraction & 0.4574 & 0.6037 & 0.5030 & 0.4316 & 0.4989 \\
No class balance & 0.4575 & \textbf{0.6114} & 0.4965 & 0.4236 & 0.4972 \\
No annealing & 0.4604 & 0.5974 & \textbf{0.5079} & 0.4204 & 0.4965 \\
Fixed acquisition batch 50 & 0.4551 & 0.6024 & 0.4818 & 0.4112 & 0.4876 \\
No DPP batch selection & 0.4634 & 0.5413 & 0.4813 & 0.3697 & 0.4639 \\
\bottomrule
\end{tabular}
\caption{Per-dataset AULC for the final method and component ablations. The mean column averages the four development datasets.}
\label{tab:ablation}
\end{table}

\begin{figure}[t]
  \centering
  \includegraphics[width=\columnwidth]{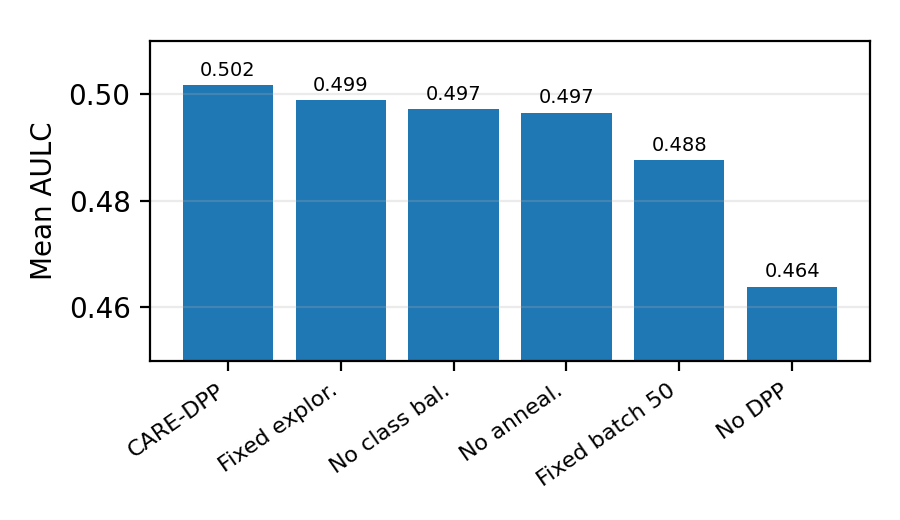}
  \caption{Mean AULC of CARE-DPP and component ablations.}
  \label{fig:ablation}
\end{figure}

\section{Discussion}
The ablation study shows that DPP batch selection is the most important component of \method. Removing it reduces mean AULC from 0.50 to 0.46, with particularly large losses on HSN and UHH (Table~\ref{tab:ablation}). This is likely because these datasets contain more heterogeneous or sparsely covered embedding regions, so selecting a non-redundant batch is especially valuable. The fixed-batch ablation also performs worse on all datasets, suggesting that smaller early acquisitions probably help because the model is updated more frequently when each new annotation has high marginal value.

The remaining ablations suggest that the best balance between uncertainty and coverage depends on the dataset. Removing annealing improves POW but hurts HSN and UHH, which could indicate that POW benefits from uncertainty earlier, whereas HSN and UHH require stronger early geometric coverage. Similarly, removing class balancing slightly improves HSN but reduces ATBFL, POW and UHH, suggesting that rare-class reweighting is useful overall but may not be uniformly optimal. The fixed-exploration ablation is close to the full method and slightly improves POW, while reducing ATBFL, HSN and UHH. This pattern suggests that stronger early random exploration likely helps broaden candidate coverage, but may occasionally over-explore when the quality score is already informative.

In our analyses, additional candidate-pool variants were explored but did not yield robust improvements over the configuration reported here. Because these analyses are not reported in detail in this paper, further research should investigate whether dataset-adaptive exploration schedules or objective-specific weight tuning can improve performance without introducing dataset-specific rules.

\section{Conclusion}
We introduced \method, a general active-learning sampler combining class-balanced uncertainty, annealed embedding novelty, strong early candidate exploration, and DPP batch diversification. The method reaches a mean development AULC of 0.50 across terrestrial and marine bioacoustic datasets. Ablations show that DPP selection and adaptive acquisition batches provide the largest gains, while class balancing, annealing and exploration scheduling provide complementary improvements. The method is submitted with reproducible BaseAL code and five-repeat result exports. Future work could further optimize the method’s hyperparameters and investigate how dataset properties, such as label density, class imbalance, and embedding-space structure, modulate the effectiveness of its individual components.

\section{Acknowledgment}
The author thanks the BioDCASE 2026 Task 4 organizers for providing the BaseAL framework and curated datasets.

\bibliographystyle{IEEEtran}
\bibliography{refs}

\end{sloppy}
\end{document}